# Highly thermally stable sub-20nm magnetic random-access memory based on perpendicular shape anisotropy

N. Perrissin[1], S. Lequeux[1], N. Strelkov[1,2], L. Vila[1], L. Buda-Prejbeanu[1], S. Auffret[1], R.C. Sousa[1], I.L. Prejbeanu[1], B. Dieny[1]

(1) Univ. Grenoble Alpes, CEA, CNRS, Grenoble INP*, INAC-SPINTEC, 38000 Grenoble, France
(2) Department of Physics, Lomonosov Moscow State University, Moscow 119991, Russia
*Institute of Engineering Univ. Grenoble Alpes

Contact email: bernard.dieny@cea.fr

## Abstract

A new approach to increase the downsize scalability of perpendicular STT-MRAM is presented. It consists in significantly increasing the thickness of the storage layer in out-of-plane magnetized tunnel junctions (pMTJ) as compared to conventional pMTJ in order to induce a perpendicular shape anisotropy (PSA) in this layer. This PSA is obtained by depositing a thick ferromagnetic (FM) layer on top of an MgO/FeCoB based magnetic tunnel junction (MTJ) so that the thickness of the storage layer becomes of the order or larger than the diameter of the MTJ pillar. In contrast to conventional spin transfer torque magnetic random access memory (STT-MRAM) wherein the demagnetizing energy opposes the interfacial perpendicular magnetic anisotropy (iPMA), in these novel memory cells, both PSA and iPMA contributions favor out-of-plane orientation of the storage layer magnetization. Using thicker storage layers in these PSA-STT-MRAM has several advantages. Thanks to the PSA, very high and easily tunable thermal stability factors can be achieved, even down to sub-10 nm diameters. Moreover, low damping material can be used for the thick FM material thus leading to a reduction of the write current. The paper describes this new PSA-STT-MRAM concept, practical realization of such memory arrays, magnetic characterization demonstrating thermal stability factor above 200 for MTJs as small as 8nm in diameter and possibility to maintain thermal stability factor above 60 down to 4nm diameter.

**Keywords**: Spin transfer torque, magnetic random access memory, perpendicular anisotropy, shape anisotropy, downsize scalability

# Main

## Introduction

A magnetic random-access memory (MRAM) is a non-volatile memory wherein one bit of information is stored by the magnetic state of a ferromagnetic (FM) layer. Microelectronic industry has recently shown a strong interest for MRAM as they are very promising for embedded RAM applications and particularly embedded FLASH replacement [1–6]. Nowadays most of the development are focused on MRAM based on out-of-plane magnetized tunnel junctions written by spin transfer torque (STT) [7–12]. The p-MTJ essentially consists of an MgO tunnel barrier (1-1.5 nm thick) sandwiched by two thin perpendicularly magnetized FeCoB layers (1-2.5 nm thick), namely the reference and storage layer. The magnetization of the reference is pinned in one specific direction by exchange coupling with a synthetic antiferromagnet (SAF). The magnetization of the storage layer can be switched between the up and down states, respectively coding the "0" and "1" of the binary logic. The state of the cell is read thanks to the tunneling magnetoresistance effect (TMR). A parallel (P) relative configuration between the magnetization of the reference and the storage layer leads to a low resistance state while an antiparallel one, to a high resistance state. TMR of up to 240 % at room temperature are nowadays obtained in highly optimized MTJs deposited in state of the art sputtering deposition tools. The energy $E_B$ required to switch the memory between these two states at temperature T is characterized by the dimensionless thermal stability factor $\Delta_T$. In macrospin approximation, $\Delta_T$ is given by eq. 1.

$$\Delta_T = \frac{E_B}{k_B T} = \frac{A}{k_B T}\left[\frac{1}{2}\mu_0 M_S^2 t(N_{xx} - N_{zz}) + K_s + K_u t\right] \quad \text{(Eq. 1)}$$

In this expression, $k_B$ and $\mu_0$ are respectively the Boltzmann constant and the vacuum magnetic permeability and T is the absolute temperature. The other parameters account for the magnetic properties of the cylindrical storage layer. $M_S$ is the saturation magnetization, $N_{xx}$ and $N_{zz}$ are respectively the in-plane and out-of plane demagnetizing factors where z refers to the out-of-plane direction, $A = \pi(D/2)^2$ is the storage layer area with D its diameter, t its thickness, $K_S$ is the interfacial anisotropy at the MgO/FeCoB interface and $K_u$ accounts for a possible uniaxial magnetocristalline or magnetoelastic anisotropy. For standard FeCoB/MgO/FeCoB p-MTJs, $N_{xx} - N_{zz} = -1$ as the thickness (1.4 nm) is much lower than the diameter (> 20 nm) leading to a negative demagnetizing anisotropy. On the contrary, the strong positive interfacial anisotropy $K_S$ ($K_S^{FeCoB}$ ~1.4 mJ/m$^2$ for Fe rich alloys)[13] acts against the demagnetizing anisotropy to pull the magnetization out-of-plane. Concerning the bulk anisotropy $K_u$, it is usually negligible in conventional p-MTJ. To fulfill industrial needs, $\Delta_T$ should be typically in the range 60 to 100 at room temperature (T=300K)[13] depending on the memory capacity and allowed bit error rate. $\Delta_T$ first scales linearly with the cell diameter for diameters larger than the exchange length (domain wall model), then quadratically for smaller diameters (macrospin model)[14,15]. Consequently, at sub-20 nm diameters, $\Delta_{300}$ inevitably drops below 60, thus limiting the downsize scalability of p-STT-MRAM.

The switching between the P and AP states is driven by STT. The critical current $I_{c0}$ is the current for which the STT compensates the Gilbert damping in the LLG equation. It is therefore the theoretical minimum current capable of writing the memory at zero temperature. For out-of-plane magnetized nanostructure, in macrospin approximation, $I_{c0}$ scales with $\Delta_T$:

$$I_{c0} = \frac{4e}{\hbar} \frac{\alpha \, k_B T}{\eta} \Delta_T \qquad \text{(Eq. 2)}$$

where $|e|$ is the absolute charge of the electron, $\hbar$ is the reduced Planck constant, $\alpha$ is the Gilbert damping and $\eta$ is the spin polarization of the current. The ratio $\Delta_T / I_{c0}$ has been proposed to be a figure of merit of STT-MRAM [16]. When the thickness of the storage layer is reduced and becomes comparable or less than the spin diffusion length, $\alpha$ increases because of the spin pumping effect [17–19], leading to an increase of $I_{c0}$. In addition, at very small thickness (typically below 1.5nm), the TMR amplitude decreases due to increased thermal fluctuations in the storage layer magnetization. Consequently, a tradeoff has to be found in conventional STT-MRAM between large thickness (i.e. 2.5 nm) for which $\alpha$ is reduced and TMR increased but effective anisotropy is no longer out-of-plane, and small thickness (i.e.~1 nm) for which the effective anisotropy is strongly out-of-plane but the damping is excessively large and the TMR weak. The tradeoff is usually found for storage layer thickness of the order of 1.4 to 1.6nm.

This paper describes a new approach which aims at overcoming what limits the downsize scalability of traditional p-STT-MRAM, namely (i) the negative contribution of the shape anisotropy and (ii) the reduction of $I_{c0}$ due to the reduction of $\alpha$ at small thicknesses. It takes advantage of a perpendicular shape anisotropy (PSA) provided to the storage layer by depositing a thick ferromagnetic (FM) layer on top of the traditional MgO/FeCoB based MTJ.

## Concept of PSA-STT-MRAM

A Perpendicular Shape Anisotropy (PSA)-STT-MRAM is an MRAM in which the thickness of the storage layer is comparable to its diameter, leading to a positive demagnetizing anisotropy contribution. In such system, all sources of anisotropy are thus favoring the perpendicular direction and no longer fight against each other. A thick ferromagnetic (FM) layer can be deposited on top of a conventional MgO/FeCoB based MTJ to keep good interfacial properties (TMR, interfacial anisotropy). All following formulas are established in macrospin approximation, which we justify in the next section. When the storage layer is made of two different layers, for example one thin FeCoB for interfacial properties of thickness $t^{FeCoB}$ and one thick FM of thickness $t^{FM}$ for providing the positive shape anisotropy, any material parameter X (X = $\alpha$, $M_S$, K) referring to the whole storage layer is given by eq. 3, where $X^{FeCoB}$ and $X^{FM}$ respectively represents the parameter X of the FeCoB and FM layer. Moreover, in order to work with analytical expressions, an approximate expression of the demagnetizing factors is used (eq. 4)[20], where $\rho = t/D$ represents the aspect ratio of the storage layer ($t = t^{FM} + t^{FeCoB}$). Under these assumptions, the stability factor $\Delta_T$ from eq. 1 is thus described by eq. 5, which is used to plot the diagram presented in fig. 1. It represents the thermal stability factor at 300K in color code for a stack FeCoB(1.4 nm)/Co(t-1.4 nm) versus diameter and total thickness t of the storage layer. The smallest thicknesses correspond to the conventional p-STT-MRAM regime (t < 2 nm) while the larger ones correspond to the novel PSA-STT-MRAM regime (t > 3 nm). The diagram clearly illustrates the inability of p-STT-MRAM to maintain $\Delta_{300} > 60$ at sub-20 nm nodes. On the other hand, once a thick FM (Co) layer is added on top of the storage layer, a strong perpendicular anisotropy can be recovered coming from the shape itself. By changing the thickness and/or the diameter of the storage layer, one can tune the stability over a very wide range, even for sub-10 nm nodes. For this example, a stability of $\Delta_{300} = 60$ can be achieved down to a diameter of 4 nm with a total storage layer thickness of 32 nm.

$$X = \frac{X^{FeCoB} t^{FeCoB} + X^{FM} t^{FM}}{t} \quad \text{(Eq. 3)}$$

$$N_{zz} = \frac{1}{1 + 4\rho/\sqrt{\pi}} = 1 - 2N_{xx} \quad \text{(Eq. 4)}$$

$$\Delta_T = \frac{\pi D^2}{4k_B T} \left[ \frac{\mu_0 M_S^2}{4} t \left(1 - \frac{3}{1 + 4\rho/\sqrt{\pi}}\right) + K_u t + K_S \right] \quad \text{(Eq. 5)}$$

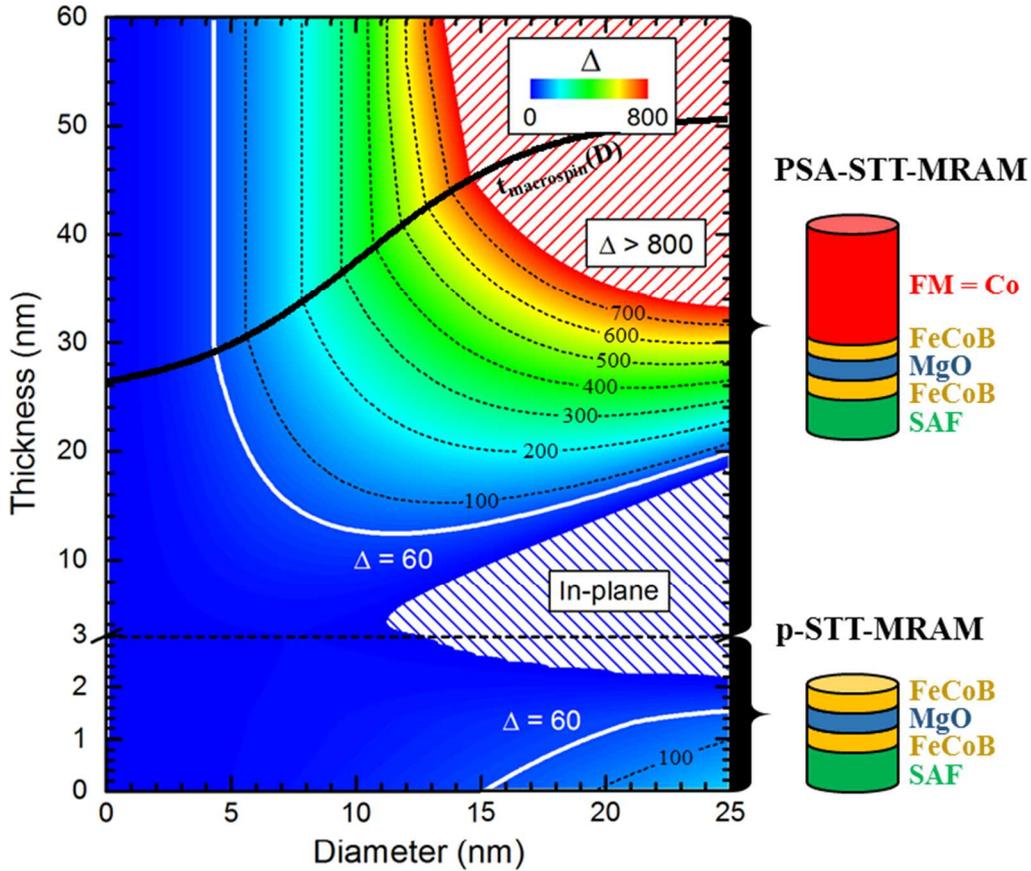

*Fig. 1. Stability diagram of a cylindrical storage layer made of FeCoB(1.4 nm)/Co(t-1.4 nm) versus its total thickness (t) and diameter (D), at room temperature (300 K). Below a total thickness (t) of 1.4 nm, the storage layer is supposed to only consist of FeCoB with $M_S^{FeCoB}$ = 1.0 10⁶ A/m and $K_S^{FeCoB}$ = 1.4 mJ/m². Above 1.4 nm, the storage layer consists of FeCoB(1.4 nm) /Co(t-1.4 nm), with $M_S^{Co}$ = 1.446 10⁶ A/m [21] and $K_u^{Co}$ = 0 J/m³. The vertical axis is cut in two parts to better show the first 3 nm, corresponding to the standard p-STT-MRAM regime. The blue dashed area represents the in-plane regime, which is off topic. The red dashed area represents the region where $\Delta_{300}$ is greater than 800. The solid black line represents the limit above which the switching behavior is no longer macrospin. Above this line, $\Delta_{300}$ is calculated via micromagnetic simulations (see fig. 2). The iso-$\Delta$ line, $\Delta_{300} = 60$, is highlighted in bold white, other iso-$\Delta$ lines are shown in dashed black lines. The smaller thicknesses correspond to*

*the standard p-STT-MRAM regime while the larger ones correspond to the PSA-STT-MRAM regime, as described by the sketches on the right of the diagram.*

## Range of validity of the macrospin approximation

Micromagnetic stable states (SS) of magnetized cylinders have been widely studied [22]. In particular, the SS of thin cobalt cylinders (thickness<diameter) with sub-25 nm diameters, corresponding to the diameter range of Fig. 1, are well described by a uniform magnetization (macrospin approximation). In order to determine the range of validity of the macrospin model during the switching process, we computed the minimum energy path (MEP) that the magnetization has to follow to switch between its two stable states (magnetization up or down), using the string theory method [23–27]. The corresponding thermal stability factors are plotted in Fig.2a. For every diameter (D<25nm), the switching of the magnetization of thin cylinders is perfectly described by a fully uniform switching (Fig. 2.b), with a stability factor matching perfectly the analytical expression given in eq. 5. For thicker cylinders, the switching mechanism is described by the nucleation of a domain wall (DW) at one end, propagation of the DW and then annihilation at the other end (Fig. 2.c). From the thickness where it becomes more energy-efficient to create a DW to the $t \to \infty$ limit, the stability factor first slowly increases before reaching an asymptotic value $\Delta^\infty$ given by the following expression (see supplemental information for more details):

$$\Delta_T^\infty = \frac{\mu_0 M_S^2}{2k_B T} \frac{\pi D^2}{4} \left(\frac{D}{2} + L_{DW}^\infty + \frac{2(L_{DW}^\infty)^2}{D + 2L_{DW}^\infty}\right), \qquad L_{DW}^\infty = \sqrt{\frac{4A_{ex}}{\mu_0 M_S^2}} \qquad \text{(Eq. 6)}$$

where $L_{DW}^\infty$ is the width of the DW and $A_{ex}$ is the exchange stiffness constant of the FM layer ($A_{ex}^{Co} \simeq 30$ pJ/m). The frontier between the two mechanisms (coherent rotation versus nucleation/propagation of DW) appears when their corresponding maximum energy states along the minimum energy path have the same energy. According to MEP simulation, $K_S$ has no noticeable influence neither on the position of the frontier between the uniform and DW regimes, nor on the value of $\Delta$ when the switching occurs by nucleation/propagation of DW (inset of Fig. 2.a). Nonetheless, the presence of an interfacial anisotropy does induce an asymmetry in the energy path between the two SS. The thickness $t_{macrospin}$ at which the switching mechanism is no longer uniform depends on the diameter. In the limit case where D → 0, one can equalize eq. 5 and eq. 6 to find that $t_{macrospin}(D = 0) = 4\frac{\mu_0 M_S^2 L_{DW}^\infty - K_S}{\mu_0 M_S^2 + 4K_u} \simeq 4L_{DW}^\infty$. The results of these simulations are used to complete Fig. 1 by adding a zone where the switching is non-uniform (above the solid black line). For the stack used in Fig. 1, it is possible to keep $\Delta_T > 60$ while maintaining a macrospin regime for diameters as small as 4 nm.

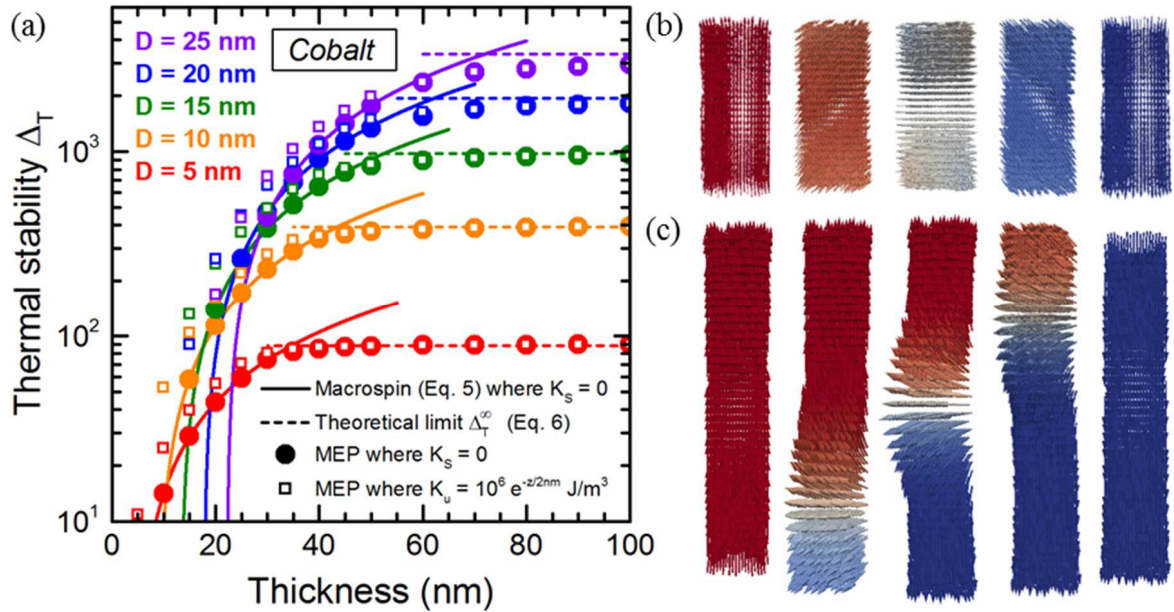

*Fig. 2.* *(a) Thermal stability factor (in logarithmic scale) versus thickness of Co cylindrical storage layer, for several diameters (color-code). The filled circles and hollowed squares symbols are extracted from micromagnetic simulations (MEP method), respectively without and with a surface anisotropy. For the latter case, the surface anisotropy is implemented as a bulk anisotropy $K_u = 10^6 \exp(-z/2nm)$ J/m³, with an amplitude which is exponentially decreasing with the depth z over a characteristic length of 2nm. The solid lines show the value of $\Delta_T$ calculated from the analytical macrospin model (eq. 5), for the case of $K_S = 0$. The dashed lines show the value of $\Delta_T^\infty$ (eq. 6). (b) and (c) respectively show a series of snapshots along the MEP illustrating a uniform switching (D = 15 nm, t = 35 nm) and a switching with nucleation/propagation of DW (D = 20 nm, t = 80 nm) for cobalt cylindrical storage layer without surface anisotropy.*

## Fabrication

The following stacks were investigated experimentally, with all thicknesses given in nm: SiO$_2$/Pt(25)/SAF/Ta(0.3)/FeCoB(1.1)/MgO(1.2)/FeCoB(1.4)/W(0.2)/FM(t$^{FM}$)/Ta(1)/Ru(10)/Ta(150). The SAF is made of two Co/Pt multilayers spaced by a 0.8 nm-thick Ru layer. Going from bottom to top of the stack: Pt(25) constitutes the bottom electrode, SAF/Ta(0.3)/FeCoB(1.1)/MgO(1.2)/ FeCoB(1.4) is a standard p-MTJ, FM(t$^{FM}$) is the thick part of the storage layer added to get the PSA (FM = Co or NiFe) with t$^{FM}$ fixed at 60 nm in this study, and finally Ta(150) is a hard mask for the etching. FeCoB stands for Fe$_{64}$Co$_{16}$B$_{20}$. The W(0.2) layer has two purposes: it absorbs the B away from the FeCoB during the annealing process and it makes a structural transition between the bcc part of the stack next to the MgO barrier and the fcc parts of the stack in the SAF and FM layer. Nevertheless, this layer is so thin that FeCoB(1.4) and FM(t$^{FM}$) are still coupled by exchange and they can be considered as one continuous magnetic layer. The fabrication process is illustrated in Fig. 3 and briefly described below. The Ta(150) hard mask pillar is first defined by e-beam lithography then etched by RIE. The MTJ stack is then etched by IBE in three steps. (#1) The storage layer is first etched at normal incidence (90°). This yields lot of redeposition on the pillars sidewalls, leading to an increase of the effective diameter. (#2) From the MgO layer to the bottom electrode, the stack is etched at intermediate angle (30°) to avoid any shorts due to redeposition on the sides of the tunnel barrier. (#3) Patterned cylinders are finally trimmed at grazing angle (10°). It has been found that a more energetic trimming beam leads to better devices. This process can produce sub-10 nm cylinders with a very high aspect ratio.

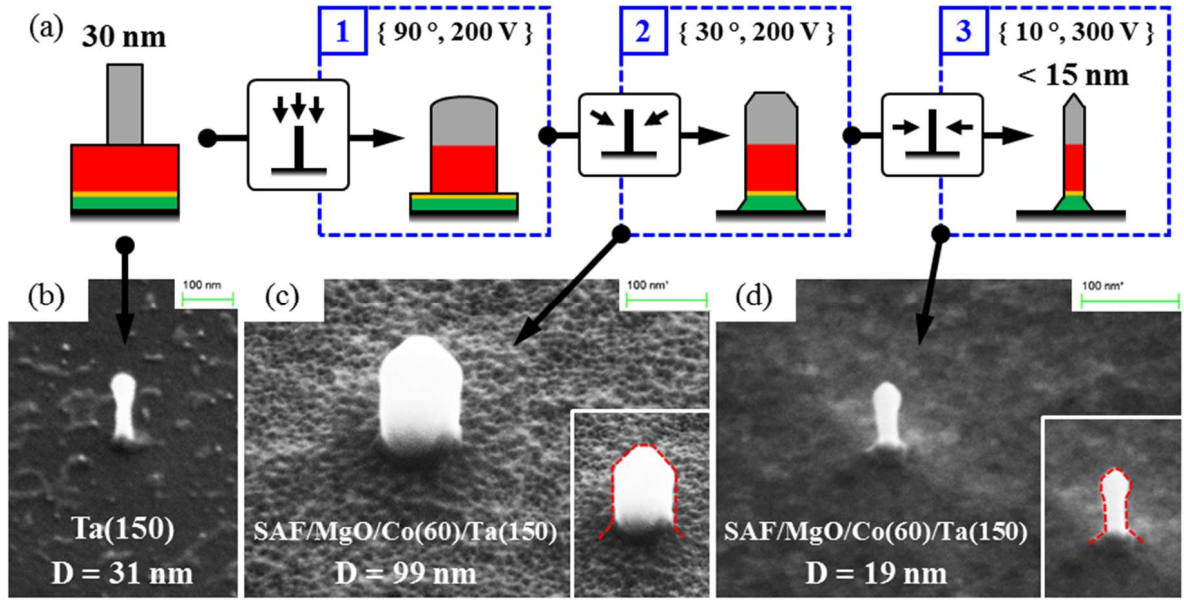

*Fig. 3.* (a) Series of sketches of the etching process. The color grey, red, yellow and green respectively represent Ta, FM, MgO and SAF. Black arrows link a SEM image to its corresponding sketch. (b) SEM image of Ta(150) with D = 30 nm, observed at 45°. (c) SEM image of SAF/MgO/Co(60)/Ta(150) with D = 99 nm after step #2, observed at 45°. In inset, we highlight with a red dashed line the edges of the pillar to better see its shape. (d) SEM image of SAF/MgO/Co(60)/Ta(150) with D = 19 nm after step #3, observed at 45°. It illustrates the effect discussed in the method section, namely the fact that Ta is etched more slowly that FM, leading to a bigger Ta head.

## Magnetic and electrical results

In this section, we study the properties of MTJs fabricated following the process illustrated in Fig.3, where two different materials were used for the storage layer of thickness $t^{FM}$=60 nm: NiFe and Co. In Fig 4.a and 4.b (respectively NiFe and Co storage layer), the evolutions of the resistance R in response to perpendicular field H are shown for MTJs of various D. The diameter of each MTJ, varying in both cases from around 8 nm to 15 nm, are determined electrically from the parallel state resistance value and the RA product (=12.9 $\Omega.\mu m^2$ and 13.5 $\Omega.\mu m^2$ for NiFe and Co respectively). The observation of a square hysteresis loop indicates in both cases a perpendicular easy axis mainly due to the shape anisotropy. The decreasing values of coercive field Hc with increasing diameter D also provide good evidence that the shape anisotropy is at the origin of the energy barrier between the two ground states of out-of-plane magnetization. The trend of Hc, decreasing as D increases, can also be seen on Fig 4.c, which shows the distribution of Hc as a function of the diameter D for NiFe (black circles) and Co (red circles) storage layers. In the case of a NiFe storage layer, the coercive field Hc varies from 1260 Oe for the smallest diameter achieved at 8.3 nm to several hundred Oe for the largest diameters around 30 nm. For the Co storage layer, coercive fields exhibit higher values, varying from 2500 Oe to 600 Oe for corresponding diameters between 8.1 and 30 nm. The different values of Hc, lower in the case of a NiFe than of a Co storage layer, is explained by the lowest saturation magnetization $M_s$ for NiFe in comparison to Co ($Ms_{NiFe}$=800emu/cm$^3$ vs $Ms_{Co}$=1420emu/cm$^3$ at RT). Finally, Fig 4.d represents the distribution of the TMR ratio measured on NiFe and Co based storage layer (respectively black and red circles). As the TMR depends on the interfacial properties next to the MgO barrier, there is no expected correlation between TMR and diameter. Nevertheless, a larger averaged TMR ratio can be noticed for the FeB/W/Co storage layer, reaching a maximum at 92 %, whereas the highest TMR for FeB/W/NiFe based MTJ reaches 60%. This difference may be ascribed to a reduced amount of thermal fluctuations at room temperature next to the MgO interface in the Co based storage layer as compared to the NiFe based one due to the higher Curie temperature of Co ($Tc_{NiFe}$=830K vs $Tc_{Co}$=1400K).

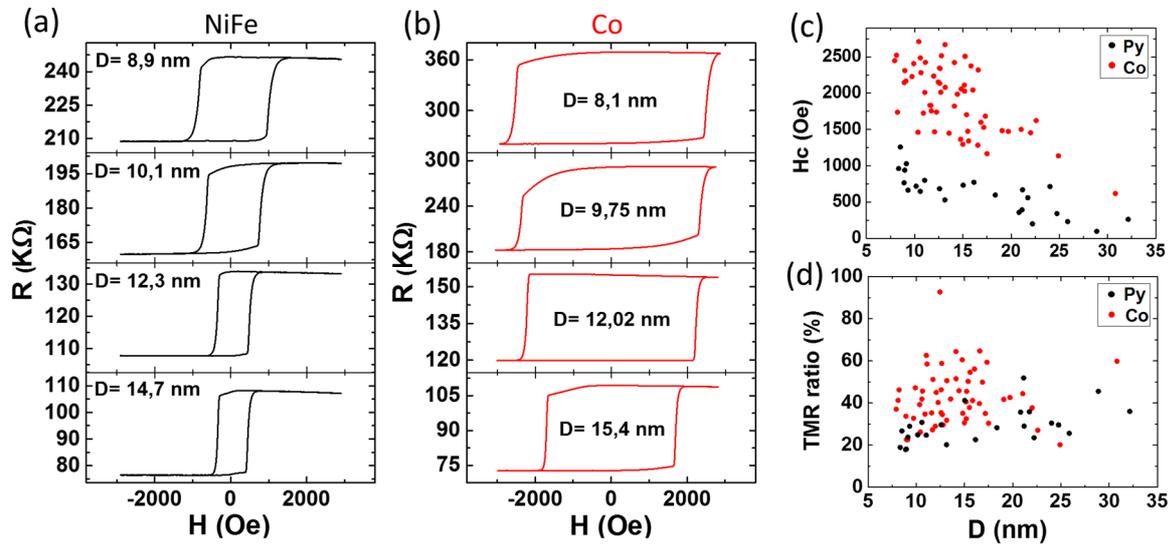

*Fig. 4.* (a), (b) Evolution of the resistance R as a function of a perpendicular field for MTJ of various diameters D in the cases of a 60 nm thick storage layer of NiFe and Co respectively. (c) Distribution of coercive fields Hc as a function of D for NiFe (black circles) and Co (red circles) storage layer. (d) Distribution of the TMR ratio as a function of D for NiFe (black circles) and Co (red circles) storage layer.

We next evaluate how small these PSA-MTJ can be made still keeping a thermal stability factor above 60. The magnetic study was conducted on MTJs with FeCoB/Co based storage layer which corresponds to the simulation results shown previously (Fig.1 and 2). The parameter Δ of our Co based PSA-STT-MRAM was experimentally determined by fitting the field dependence of $P_{sw}$ with a generalization of Sharrock model [28]. This allows to extract Δ and the anisotropy field $H_k$ at a fixed field sweep rate $R_{sweep}$ [29] (see method in annex for details). At room temperature, remarkably large thermal stability factors varying from 300 to 700 are extracted from the data, for electrical diameters from 8 nm to 13 nm. In conventional STT-

MRAM based on iPMA, it would be impossible to get such large thermal stability factors in this range of diameters.

These obtained values of Δ are in good agreement with the results of simulations shown in Fig. 1 and 2 in the case of a Co storage layer of thickness $t^{FM}$=60 nm. Nonetheless, as the method used to extract Δ from the data is based on a macrospin model [29], it is wise to discuss the validity of these values. Indeed, the macrospin approximation can yield an overestimation of Δ if the magnetization reversal actually takes place by domain wall nucleation/propagation. This can be observed in the simulations results shown in Fig. 2(a) by the difference between solid lines (macrospin) and square symbols (micromagnetic) for thicknesses larger than $t_{macrospin}$. It can also be noticed that the larger the diameter, the greater the overestimation of the thermal stability factor. Despite the overestimation at large D, for the smallest diameters achieved (down to 10 nm), the thickness $t^{FM}$=60 nm is close enough to the boundary between the two regimes (Fig. 2(a)), thus yielding only a small overestimation of the stability factor.

Finally, because of their too large values of Δ, the junctions with $t^{FM}$=60 nm could not be switched by STT. Therefore, PSA-MTJs with 12 nm thick FeCoB(2)/CoFeB(8)/FeCoB(2) free layer (with a MgO capping layer on top) were patterned. Those exhibit lower Δ values, typically between 20 and 60 depending on their diameter. As shown in Fig. 5, STT switching could then be realized between parallel and antiparallel states and vice versa at zero external field. Therefore, in order to switch PSA-MTJ of low diameter and high stability, the resistance-area (RA) product of the tunnel barrier has to be lowered since the critical voltage (eq. 2) is directly proportional to RA.

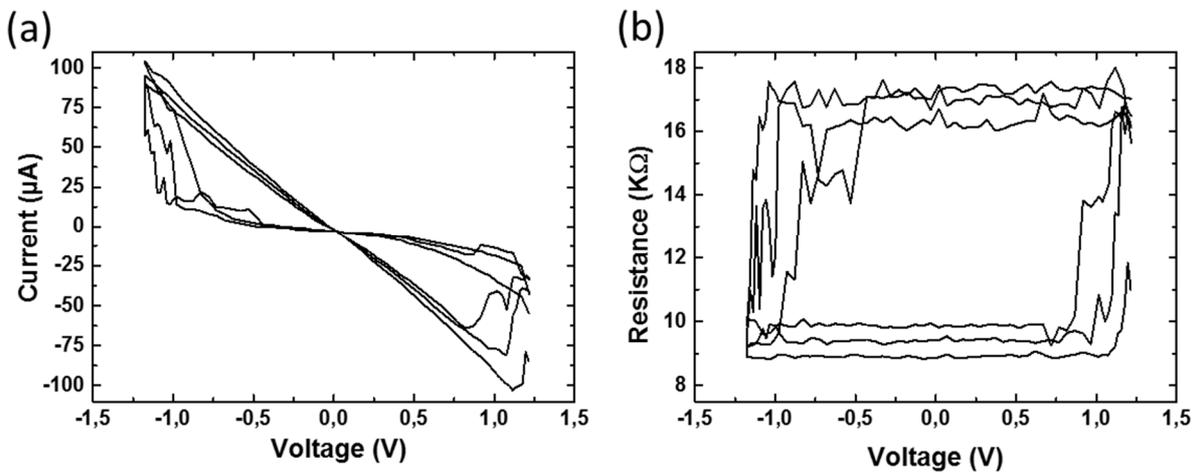

*Fig. 5. Evolution of current (a) and resistance (b) versus applied voltage at zero external field. illustrating STT switching in a PSA-MTJ with 12 nm thick FeCoB(2)/CoFeB(8)/FeCoB(2) free layer, with a MgO capping layer on top, and a diameter of 17 nm.*

The conclusion here is that PSA opens a route towards high stability factor in very small diameter pMTJ. Of course, as it is known from eq.2, for the writability of the storage layer, Δ should not be chosen too high. This is also confirmed by the results of STT switching (Fig. 5) measured on devices that have a reduced thermal stability due to the thinner free layer in comparison with PSA-MTJ of $t^{FM}$=60 nm. Values of Δ typically in the range 60 to 100 should be chosen to maintain the write voltage low enough compared to the dielectric breakdown voltage of the barrier. Similar conclusions were very recently reached by Watanabe et al[30].

## Conclusion

This paper presents a novel approach to increase the downsize scalability of conventional p-STT-MRAM by taking advantage of a perpendicular shape anisotropy. Dramatically increasing the thickness of the storage layer in these PSA-STT-MRAM offers several advantages. Firstly, the shape anisotropy no longer acts against the stability of the storage layer magnetization but constitutes a robust source of bulk perpendicular anisotropy which comes on top of the interfacial anisotropy originating from the MgO/FeCoB interfaces. For practical devices, the diameter and thickness should be chosen so that the thermal stability factor lies in the range $\Delta_{300} = 60 - 100$ depending on the application. For these combinations of parameters, the storage layer magnetic behavior is generally well described in macrospin approximation. It is only at very small nodes (around sub-4 to 8 nm depending on the $M_S$ value and exchange stiffness) that switching may start occurring by DW nucleation/propagation. In this regime, the thermal stability factor can be calculated by the MEP method. Our estimates based on experimental data and micromagnetic simulations indicate that thermal stability factor above 60 could be maintained for diameters as small as 4nm.
Secondly the use of thick storage layer allows designing storage layer such that its interfacial part in contact with the tunnel barrier provides high TMR while its bulk part provides low Gilbert damping in addition to the thermal stability.

Acknowledgements:
This work was funded by ERC Adv grant MAGICAL 669204.
The authors would like to thank Jannier Roiz for fruitful discussions and valuable suggestions.

# Methods

## Fabrication

The IBE tool used is the Scia Mill 150 from Scia Systems. The starting point of the IBE process is a 30 nm diameter cylinder of Ta(150). During the following etching, the plasma and neutralizer properties do not change, only the extraction grid voltage ($V_g$) and the angle of attack ($\beta$) do. The process is done in three steps. (#1) A first step consists in etching at normal incidence $\beta = 90°$ with a low grid voltage $V_g = 200$ V, until MgO is reached. The progression is followed by a SIMS detection. The 90° angle allows to obtain fairly vertical sidewalls. However, this inevitably comes with lot of redeposition around the pillars, leading to a significant increase of the pillar diameter. (#2) When MgO is reached, the angle of attack is set at 30° while keeping the same grid voltage, until the bottom electrode is reached. This angle is such that the lateral etching rate is slightly faster than the redeposition rate. This avoids the formation of shorts across the barrier due to redeposition of the pillar sidewalls. This angle also yields a conical shape to the reference SAF, which is not critical as long as the storage layer keeps its cylindrical shape. At the end of this step, the pillar diameter is slightly reduced by about 5nm. The low grid voltage yields a quite slow etching, which allows to well master the endpoint of each step. (#3) Finally, a last step consists in trimming the pillar to the desired diameter. The best conditions have been found for a grazing angle of $\beta = 10°$ and a higher grid voltage $V_g = 300$ V. The progression cannot be followed by SIMS detection as the signal is too weak. It is therefore performed by controlling the etching time after having calibrating the etching speed. During the trimming process, we first etch back the redeposited material, which is mainly from the storage layer. Then, once the diameter becomes smaller than the initial diameter of the Ta hard mask, different materials are etched, therefore at different speeds. In particular, the Ta is etched more slowly than the FM so the pillar is slimmer at the FM level than at the Ta level, yielding a bigger Ta head (fig. 3.d).

## Electrical measurements and extraction of thermal stability factor Δ

All the measurements were performed at room temperature. A standard electrical probing system able to apply out-of-plane ac/dc magnetic fields was used for electrical characterization of the junctions. A current source was used to apply low DC current of the order of 0.2 µA. A voltmeter was used to measure the voltage across the MTJ and derive its resistance.

In order to extract the thermal stability factor Δ, the field dependence of Psw was first determined by measuring the distribution of switching fields for both P to AP and AP to P switching on many RH loops (typically between 300 and 500 loops). Then, the field dependence of Psw was fitted using the following expression [29]:

$$\text{Psw (H)} = 1 - \exp\left[\frac{-H_K \cdot f_0 \cdot \frac{\sqrt{\pi}}{2}}{R_{sweep} \cdot \sqrt{\Delta}} \cdot erfc\left[\sqrt{\Delta} \cdot \left(1 - \frac{H}{H_K}\right)\right]\right] \quad \text{(Eq. 8)}$$

where $f_0 \sim$ 1 GHz is the attempt frequency and $erfc$ is the complementary error function. $R_{sweep}$ represents the field sweep rate, here fixed at 5.4 kOe/s. fig. S1 shows that the field dependence of Psw is well fitted by eq. 8. This allows to extract $\Delta_{P-AP}= 308 \pm 10$, $\Delta_{AP-P}= 302 \pm 10$, and anisotropy field $H_K = 2120$ Oe, for a 8.3 nm diameter MTJ.

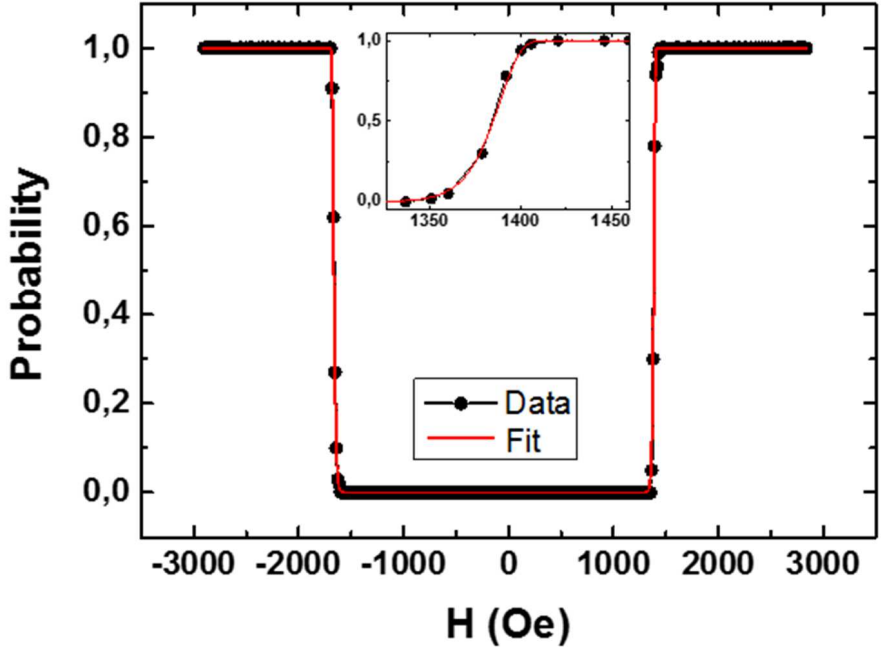

***Fig. S1.*** *Evolution of the probability of switching (Psw) as a function of the perpendicular applied field, determined from the switching field distribution obtained after 300 RH loops at a fixed field sweep rate of 5.4 kOe/s. Black line with circle symbols represents the data. Red solid line shows the fit according to eq.8. In inset is shown a zoom on the transition from P to AP states.*

## Estimation of the energy barrier for uniform and DW-based switching

To estimate the energy barrier and stability factor in the case of uniform transition, one can use a simple macrospin model. The free energy density of the ferromagnetic cylinder uniformly magnetized along one of the main axes is given by the following expression:

$$E_{\xi\,\Omega} = \frac{1}{2}\mu_0 M_s^2 N_\xi, \qquad \text{(Eq. 9)}$$

where $N_\xi$ is a demagnetizing factor of the axis $\xi = x, y, z$. We can use the approximate expressions of these factors [20]:

$$N_z = \frac{1}{2\rho + 1}, \qquad N_x = N_y = \frac{\rho}{2\rho + 1}, \qquad \text{(Eq. 10)}$$

where $\rho = t/D$ is the aspect ratio of the cylinder, $t$ and $D$ being respectively its height and diameter. In this case, the energy barrier can be obtained as the energy difference between x-magnetized metastable state and z-magnetized stable state:

$$E_B = \frac{1}{2}\mu_0 M_s^2 (N_x - N_y)\Omega = \frac{\mu_0 M_s^2}{2}\left(\frac{t-D}{2t+t}\right)\frac{\pi D^2 t}{4}, \tag{Eq. 11}$$

where $\Omega$ is the volume of the cylinder.

In case of magnetization reversal occurring by domain wall nucleation/propagation, the energy barrier can be estimated in the limit of very long heigh $t \gg D$. The distribution of the free energy density inside the magnetic wire can be written as follows:

$$E_{\text{DW }\Omega} = \frac{1}{2}\mu_0 M_s^2 (N_z \sin^2 \varphi(z) + N_x \cos^2 \varphi(z)) + A_{ex} \cdot (\varphi'(z))^2,$$

where $\varphi(z)$ is the angle between the magnetization at point $z$ and the x-axis, $A_{ex}$ is an exchange constant of the considered material. By using the well-known expression for $\varphi(z)$ profile of a domain wall:

$$\varphi(z) = -\frac{\pi}{2} + 2 \text{ atan } e^{-z/t_{DW}},$$
$$t_{DW} = \sqrt{\frac{2A_{ex}}{\mu_0 M_s^2 (N_x - N_z)}} \tag{Eq. 12}$$

where $t_{DW}$ – is the domain wall effective length, and after integrating the resulting $E_{\text{DW }\Omega}$ for $z$ from $-t/2$ to $t/2$, an averaged energy density expression is obtained.

$$E_{\text{DW }\Omega} = \frac{\mu_0 M_s^2}{2} N_z + \frac{2A_{ex} + t_{DW}^2 \mu_0 M_s^2 (N_x - N_z)}{t_{DW} t} \tanh \frac{t}{2 t_{DW}} \tag{Eq. 13}$$

Nevertheless, eq. 13 doesn't fit properly the numerical results from MEP simulations. Consequently and to have a better agreement with the simulations, we propose to slightly modify this expression by the following. The system can be considered as formed of three parts: one cylinder of thickness $t_{DW}$ containing the DW (where the magnetization goes from *down* to *up*) sandwiches by two uniformly magnetized cylinders (one *down*, one *up*) of thickness $t/2$. With this picture in mind, the first term of eq. 13 can be viewed as the energy density of the uniformly magnetized cylinders and the second term as the energy density of the DW. We therefore replace eq. 13 by eq. 14, where $N_{z\,1/2}$ is the demagnetizing factor along the z direction of a cylinder with a thickness equal to $t/2$, and where $N_{x\,DW}$ is the demagnetizing factor along the x direction of a cylinder with a thickness equal to $t_{DW}$.

$$E_{\text{DW }\Omega} \equiv \frac{\mu_0 M_s^2}{2} N_{z\,1/2} + \frac{2A_{ex} + t_{DW}^2 \mu_0 M_s^2 N_{x\,DW}}{t_{DW} t} \tanh \frac{t}{2 t_{DW}} \tag{Eq. 14}$$

The energy barrier is then simply given by $E_B = (E_{\text{DW }\Omega} - E_{z\,\Omega})\Omega$. In the limit of infinite height, the energy barrier reads as given in eq. 6.